\documentclass[prl,superscriptaddress,showpacs,twocolumn]{revtex4-1}

\usepackage{graphicx}
\usepackage{color}
\usepackage{verbatim}
\usepackage{subfigure}
\usepackage{amssymb}
\usepackage{amsmath}
\usepackage[normalem]{ulem}

\begin{document}
	
\title{Solvation in Space--Time: Pre-transition Effects in Trajectory Space}
\author{Shachi Katira}
\affiliation{
Department of Chemical and Biomolecular Engineering, University of California, Berkeley, CA, USA}

\author{Juan P. Garrahan}
\affiliation{School of Physics and Astronomy, University of Nottingham, Nottingham, NG7 2RD, UK}
\affiliation{Centre for the Mathematics and Theoretical Physics of Quantum Non-Equilibrium Systems, University of Nottingham, Nottingham, NG7 2RD, UK}

\author{Kranthi K. Mandadapu}
\affiliation{
Department of Chemical and Biomolecular Engineering, University of California, Berkeley, CA, USA}
\affiliation{Chemical Sciences Division, Lawrence Berkeley National Laboratory, Berkeley, CA, USA}
\begin{abstract}
We demonstrate pre-transition effects in space--time in trajectories of systems in which the dynamics displays a first-order phase transition between distinct dynamical phases. These effects are analogous to those observed for thermodynamic first-order phase transitions, most notably the hydrophobic effect in water. Considering the (infinite temperature) East model as an elementary example, we study the properties of 
``space--time solvation'' by examining trajectories where finite space--time regions are conditioned to be inactive in an otherwise active phase. 
We find that solvating an inactive region of space--time within an active trajectory shows two regimes in the dynamical equivalent of solvation free energy: an ``entropic'' small solute regime in which uncorrelated fluctuations are sufficient to evacuate activity from the solute, and an ``energetic" large solute regime which involves the formation of a solute-induced inactive domain with an associated active--inactive interface bearing a dynamical interfacial tension. We also show that 
as a result of this dynamical interfacial tension there is a dynamical analog of the hydrophobic collapse that drives the assembly of large hydrophobes in water. We discuss the general relevance of these results to the properties of dynamical fluctuations in systems with slow collective relaxation such as glass formers.  
\end{abstract}
\maketitle
\begin{figure}[t!] 
\includegraphics[width=1.0\columnwidth]{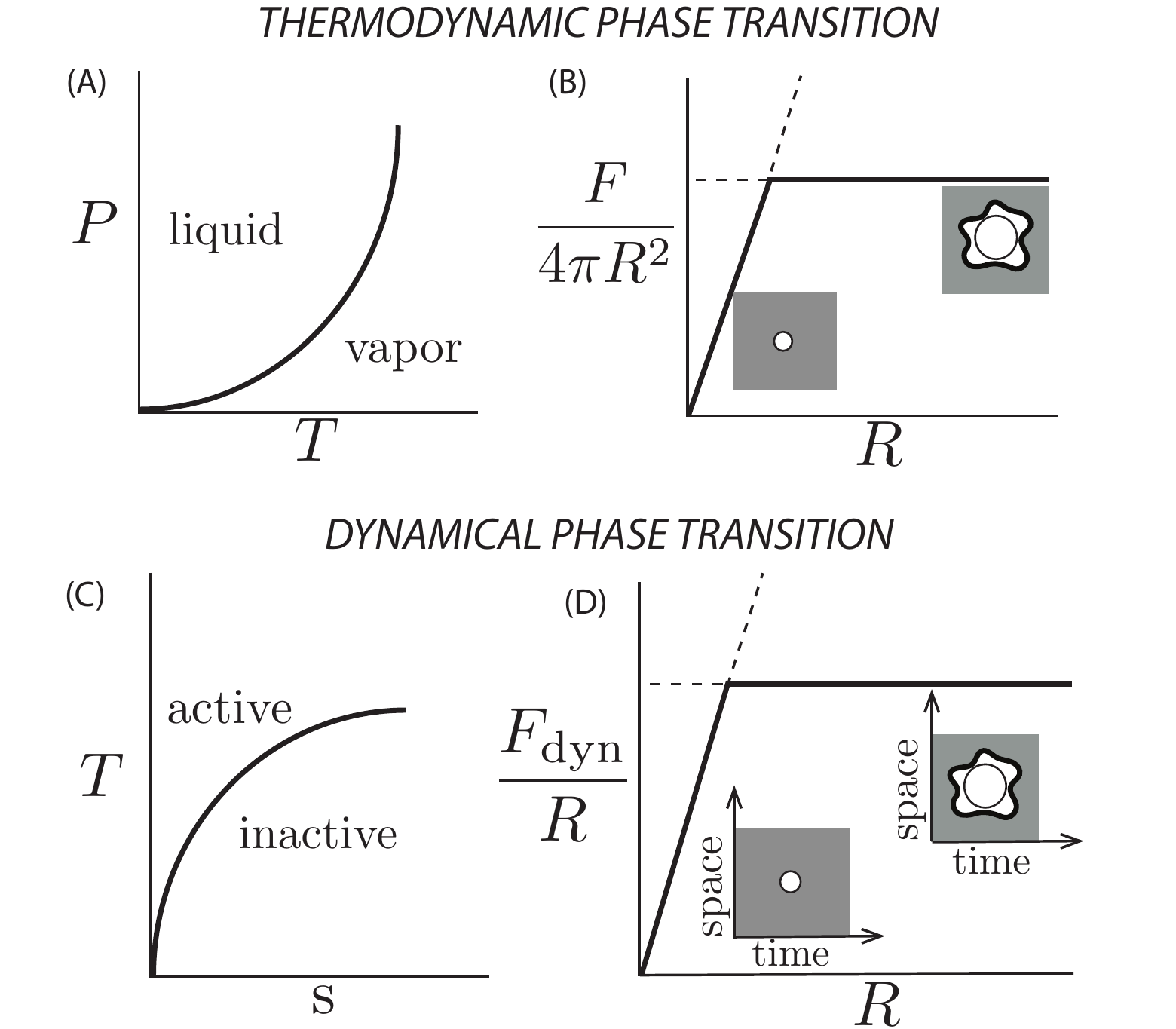}
\caption{(A) Thermodynamic first-order phase transition between liquid and vapor phases. (B) Solvation free energy $F$ per unit area of a hard sphere with radius $R$ in a liquid shows two regimes: a small solute entropic regime, and a large solute energetic regime which involves formation of a vapor domain around the sphere with an associated liquid--vapor interface; adapted from~\cite{chandler2005interfaces}. 
(C) Trajectory space phase diagram showing a dynamical first-order phase transition between active and inactive phases, cf.\ \cite{elmatad2013space}).
(D) Expected dynamical hydrophobic-like effect for ``space--time solutes''.}
\label{fig:analog}
\end{figure}

\noindent
{\bf \em Introduction.} %\textcolor{red}{A relevant surface or solute, when embedded in a bulk phase, can induce a thermodynamically stable domain that resembles the other phase across a first-order phase transition \cite{lipowsky1984surface,lipowsky1982critical,lum1999hydrophobicity,chandler2005interfaces,katira2016pre}. We refer to this effect as a \emph{pre}-transition effect because it occurs in regions of the phase diagram proximal to a first-order phase transition.}\, 
In the presence of a relevant {surface} \cite{lipowsky1984surface,lipowsky1982critical} or solute~\cite{lum1999hydrophobicity,chandler2005interfaces,katira2016pre}, systems that exhibit thermodynamic first-order phase transitions manifest pre-transition effects, which occur in the vicinity of the phase transition. 
The most notable example of a solute-induced pre-transition effect is the {\em hydrophobic effect} in water, which is related to its first-order liquid--vapor phase transition~\cite{lum1999hydrophobicity,huang2000cavity,huang2001scaling}  (Figs.~\ref{fig:analog}A,\,B). To solvate a small ($<$\,1\,nm) region of space constrained to contain no water molecules (i.e., a hard sphere), the free energy of solvation is entropic in origin and scales as the volume of the sphere (Fig.~\ref{fig:analog}B). However, for a larger region ($>$\,1\,nm) it is thermodynamically favorable to create a domain resembling the vapor phase around the hard sphere, and the free energy of solvation scales as the surface area of the sphere (Fig.~\ref{fig:analog}B). The interface between this induced vapor domain and the bulk liquid resembles the liquid--vapor interface at coexistence~\cite{chandler2005interfaces,mittal2008static}. Two such hydrophobic spheres assemble to reduce the interfacial energy between the induced vapor domains and the bulk liquid, termed {\em hydrophobic collapse}~\cite{willard2008role}. Analogous pre-transition effects have been shown to result in assembly of transmembrane proteins in model lipid bilayers with first-order order--disorder phase transitions~\cite{katira2016pre}.
\begin{figure*}[ht] %  figure placement: here, top, bottom, or page
\includegraphics[width=\textwidth]{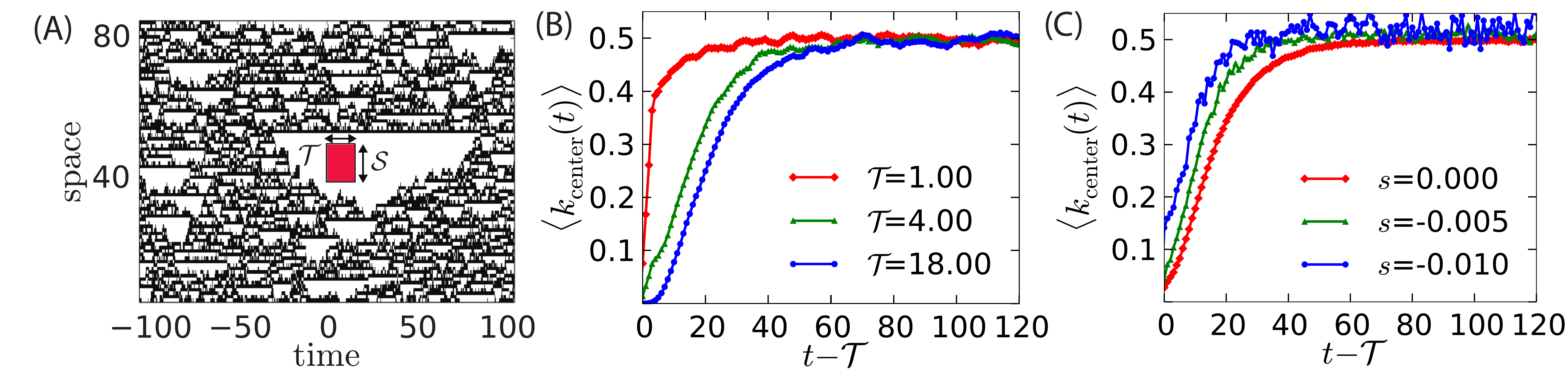}
\caption{Solutes in trajectory space: (A) An East model trajectory at $T=\infty$ containing a solute (red).  Sites with $n=0$ ($n=1$) are in white (black). The solute has spatial size $\mathcal{S}$ and temporal size $\mathcal{T}$. For clarity, the solute begins at time $t=0$. (B) Average activity of the spin at the midpoint of the solute as a function of time from the edge of the solute (for $\mathcal{S}=12$ and various $\mathcal{T}$). Away from the solute, the average activity attains its bulk value of $1/2$. For large enough solutes an inactive domain is induced around the solute, as indicated by the dip near $t-\mathcal{T} = 0$. (C) Pre-transition effects persist away from the dynamical transition at $s=0$ (here $\mathcal{S} = 12$ and $\mathcal{T}=4$).
}
\label{fig:activityprofile}
\end{figure*}

In this paper we show that proximity to a dynamical phase transition results in analogous effects in the solvation of ``space--time solutes''. Dynamical first-order phase transitions are trajectory space counterparts of thermodynamic first-order phase transitions with two co-existing ensembles of trajectories exhibiting a sharp jump in the value of an order parameter. Several such dynamical phase transitions have been observed in classical~\cite{bertini2015macroscopic,bodineau2004current,bodineau2005distribution,hurtado2011spontaneous,lecomte2007thermodynamic,garrahan2007dynamical,hedges2009dynamic,limmer2014theory,vaikuntanathan2014dynamic,tizon2017order} as well as quantum~\cite{garrahan2010thermodynamics,lesanovsky2013characterization,ates2012dynamical} systems with broad applications in the description of glassy dynamics, driven systems, and active matter.  Here we demonstrate pre-transition effects in trajectory space 
%--- i.e.\ a dynamical version of thermodynamic pre-transition effects such as hydrophobicity --- 
using the East model as an example, which is a kinetically constrained model for glass formers and exhibits a dynamical first-order transition between active and inactive trajectories~\cite{garrahan2007dynamical,garrahan2009first}. We show that active trajectories conditioned on containing space--time solutes, i.e., regions of inactivity, display pre-transition phenomena that are the dynamical analog of thermodynamic pre-transition effects such as the hydrophobic effect. Although we illustrate this effect for the (infinite temperature) East model, our findings should be generic in systems with first-order dynamical transitions such as glass formers.
\vspace{0.07in}

\noindent
{\bf \em Model and active--inactive transitions.} The East model~\cite{jackle1991hierarchically} displays an active--inactive phase transition {\em at all temperatures} $T$ due to the constrained nature of its dynamics (see below). For further simplicity we consider $T=\infty$ where typical dynamics is not glassy, in contrast to low temperatures. This will further emphasize the generality of our results, {which will be more pronounced at lower temperatures where it is easier to create and sustain correlations.}

 The East model consists of a one-dimensional lattice of $N$ binary variables, $n_i = 0,1$, with $i=1, \ldots, N$, and a non-interacting energy function~\footnote{The energy function for the East model is defined as $H=\sum_i n_i$. The concentration of excited spins is given by $c=$\unexpanded{$ \langle n_i \rangle$}$=\exp(-1/k_\mathrm{B}T)/(1+\exp(-1/k_\mathrm{B}T))$, where $k_\mathrm{B}$ is the Boltzmann constant. $c$ is 1/2 at $T=\infty$}, which plays no role at $T=\infty$.  The dynamics occurs via single spin flips where sites are constrained by their left nearest neighbor: site $i$ can flip only if site $i-1$ is {\em excited}, i.e., $n_{i-1} = 1$. At $T=\infty$, if a spin is {\em facilitated}, it can flip up or down with equal rate. 

At infinite temperature the equilibrium state is a non-interacting collection of spins with equal probability of being up or down, the concentration of excitations $c$ is large, and the constraints play no significant role as most sites are facilitated. Nevertheless the system is at the brink of a dynamical singularity.
Trajectories of this system can be classified by a dynamical order parameter, for example the {\em dynamical activity} (i.e., the total number of configurations in a trajectory). The distribution of such a dynamical order parameter is strongly non-Gaussian. Specifically, if one defines 
the associated moment generating function (which plays the role of a dynamical partition sum)
\begin{equation}
\label{eq:Zst}
Z(s,t_\mathrm{obs}) = \sum_{\mathrm{traj}} P(\mathrm{traj}) e^{-sK_\mathrm{traj}} ,
\end{equation} 
where $K_\mathrm{traj}$ is the activity of a trajectory and $s$ a ``counting field'', 
one finds that $Z(s,t_\mathrm{obs})$ displays a first-order singularity at $s=0$ in the limit of long trajectories and large system sizes. This corresponds to a first-order phase transition between two dynamical phases, an active phase (for $s<0$) where trajectories exhibit finite activity, and an inactive phase (for $s>0$) where trajectories exhibit vanishing activity~\cite{garrahan2007dynamical,garrahan2009first}. The set of trajectories with probabilities proportional to the summand of Eq.~(\ref{eq:Zst}) is sometimes called the $s$-ensemble \cite{garrahan2009first}.
\vspace{0.07in}

\noindent
{\bf \em Space--time solutes.} To demonstrate pre-transition effects, we consider the ensemble of trajectories of length $t_\mathrm{obs}$ conditioned on having zero activity within a fixed region of space--time designated the {\em solute} (see Fig.~\ref{fig:activityprofile}A). The spatial and temporal sizes of the solute are denoted by $\mathcal{S}$ and $\mathcal{T}$ respectively. We define the dynamical partition sum for such an ensemble as
\begin{equation}
\label{eq:Zsolute}
Z_{\mathrm{solute}=\mathcal{S}\times\mathcal{T}}(s,t_\mathrm{obs}) = \sum_{\mathrm{traj}} P(\mathrm{traj}) e^{-sK_\mathrm{traj}} \delta(A_\mathrm{traj}),
\end{equation} 
where $\delta(A_\mathrm{traj})$ is the Dirac delta function and $A_\mathrm{traj}$ is the activity in the solute region of the trajectory. As in Eq.~(\ref{eq:Zst}), we have exponential tilt controlled by $s$, so that $ Z_{\mathrm{solute}=\mathcal{S}\times\mathcal{T}}(s,t_\mathrm{obs}) / Z(s,t_\mathrm{obs}) $ is the probability of inserting the solute in the trajectory. 

We harvest solute-containing trajectories using a path sampling procedure outlined in Ref.~\cite{jack2006space}. Fig.~\ref{fig:activityprofile}A shows a trajectory from this ensemble, cf.~Eq.~(\ref{eq:Zsolute}). As seen in the trajectory in Fig.~\ref{fig:activityprofile}A, the solute region constrained to have zero activity induces an inactive domain in its vicinity. %This is analogous to the hydrophobic effect where a vapor-like domain is created by a (large, $>$\,1\,nm) hard sphere in water~\cite{lum1999hydrophobicity,chandler2005interfaces}, and the orderphobic effect where a large protein in an ordered lipid bilayer induces a disordered domain in its vicinity~\cite{katira2016pre}.
To quantify the extent of the inactive domain around the solute region we examine the variation of the activity in the vicinity of the solute {in the temporal direction}. Fig.~\ref{fig:activityprofile}B shows the average intensive activity of the spin at the center of the solute at time $t$, $\langle k_\mathrm{center}(t)\rangle$, as a function of distance in time from the edge of the solute region for three different solute sizes. This temporal profile of the activity around a large solute ($\mathcal{T}=4$ and $18$) appears to be a sigmoid function between zero activity and the activity of the bulk active phase. This shows that the domain in the vicinity of the solute resembles the inactive phase. {The sigmoid function indicates the presence of a sharp, but fluctuating, interface connecting two distinct phases~\cite{rowlinson2013molecular}.} Fig.~\ref{fig:activityprofile}B also shows that the activity profiles change with solute size --- for the smallest solute there is no inactive domain created in its vicinity. {The activity profile in the spatial direction is unremarkable and does not appear to show a pre-transition domain at the conditions considered.}

In the case of thermodynamic pre-transition effects the size of the solute-induced domain decreases as the system moves away from the first-order phase transition because the chemical potential difference between the two phases exceeds $\mathcal{O}(k_\mathrm{B}T)$~\cite{lum1999hydrophobicity,katira2016pre}. Analogously, for the dynamical first-order transition considered here, we find that the size of the inactive domain around the solute decreases away from the first-order phase transition line ($s=0$), as seen in Fig.~\ref{fig:activityprofile}C. {We note that for finite-sized systems, phase coexistence occurs at a point $s>0$~\cite{nemoto2017finite,bodineau2012activity,bodineau2012finite}. As a result, $s=0$ is in the active part of the phase diagram and is also a convenient point at which to sample trajectories.}

We anticipate a similar, but reverse, effect on the other side of the phase transition line, where an active solute in a bulk inactive trajectory induces an active domain in its vicinity.%By sampling ensembles of trajectories away from this line at negative $s$ values, Fig.~\ref{fig:activityprofile}C shows activity profiles in the vicinity of a solute with size $\mathcal{T}=4$. As expected, the extent of the inactive domain created decreases with decreasing values of $s$.

\vspace{0.07in}

\begin{figure}[t!] 
\includegraphics[width=1.0\columnwidth]{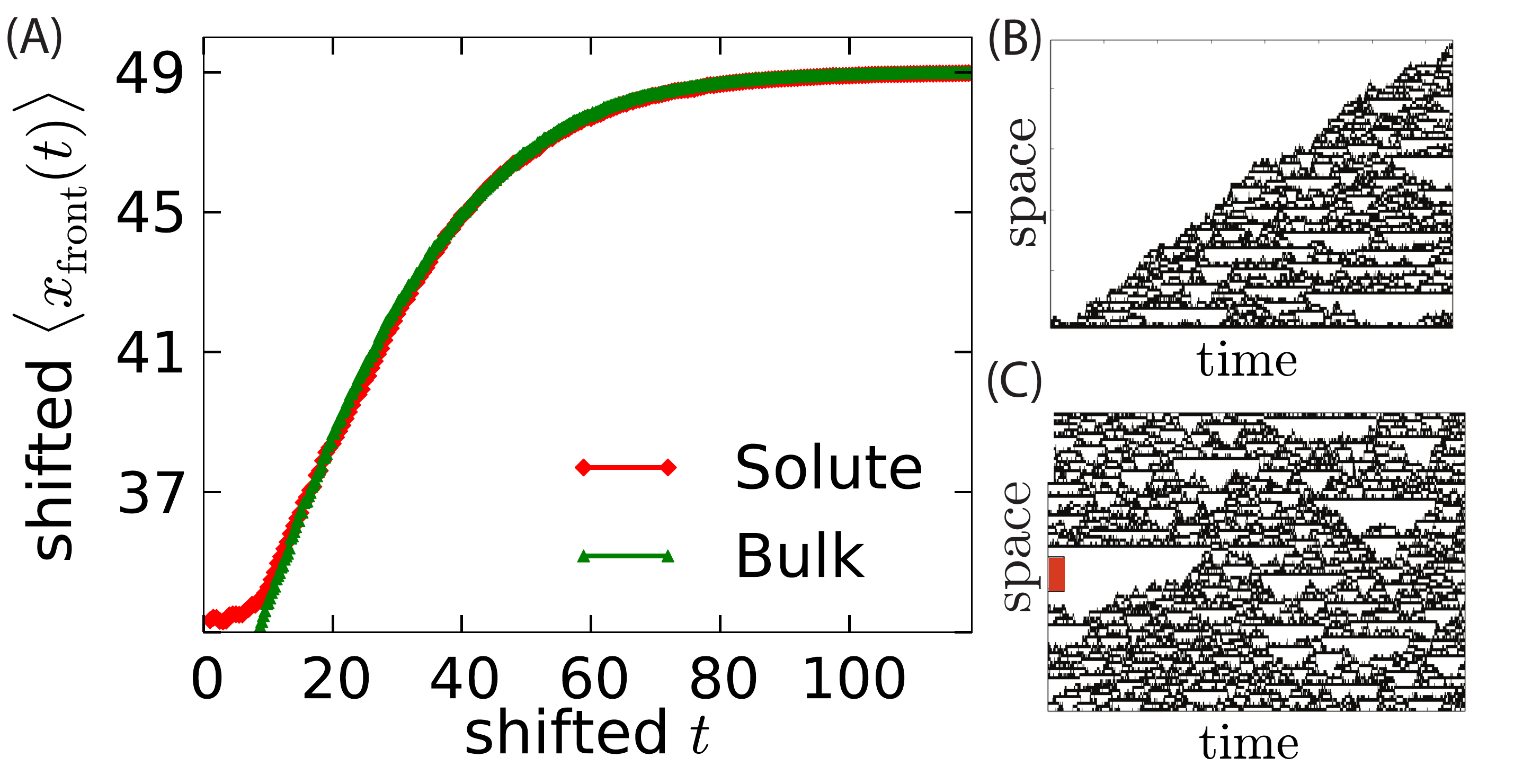}
\caption{(A) Comparison of the relative position of the interface at coexistence (B) and that induced by the solute (C). The solute-induced front is translated uniformly in space and time to coincide with the front between the bulk phases.}
\label{fig:interface}
\end{figure}

\noindent
{\bf \em Interfaces in space--time.}
The creation of an inactive domain by the solute in an otherwise active bulk phase leads to the formation of an associated interface separating the induced domain and the bulk phase. %In the case of thermodynamic phase transitions, interfaces induced by solutes away from coexistence have been shown to be quantitatively similar to interfaces between the two bulk phases at coexistence~\cite{katira2016pre,mittal2008static}. In addition to confirming the connection between solute-induced domains and the phase transition, the interfacial tension associated with these interfaces predicts solvation free energies of solutes~\cite{lum1999hydrophobicity, chandler2005interfaces}. In what follows, we demonstrate that the interfaces induced by the space--time solutes are quantitatively similar to those between the bulk active and inactive phases at coexistence. 
Fig.~\ref{fig:interface}A compares interfaces obtained at coexistence with those induced by the presence of a solute. Interfaces at coexistence are sampled from trajectories initiated with a configuration with all spins except the leftmost set to 0. This yields an ensemble of trajectories at {coexistence} %$s=0$ 
with a bulk active phase and a bulk inactive phase separated by a diagonal space--time interface as shown in Fig.~\ref{fig:interface}B. The interface can be described by the position of the front, $x_\mathrm{front}(t)$, defined as the rightmost lattice site with $n_i=1$ at time $t$~\cite{ganguly2015cutoff}. As shown in Fig.~\ref{fig:interface}A, the resulting front is in excellent agreement with the latter half of the solute-induced inactive domain in Fig.~\ref{fig:interface}C. This analysis shows that the velocity of the front, $\nu = d x_{\mathrm{front}}(t)/{d t}$ {= 0.4}, is the same in both cases. This further confirms that the domain created in the vicinity of the solute resembles the inactive dynamical phase. 

\vspace{0.07in}

\begin{figure*}[t!] 
\includegraphics[width=0.9\textwidth]{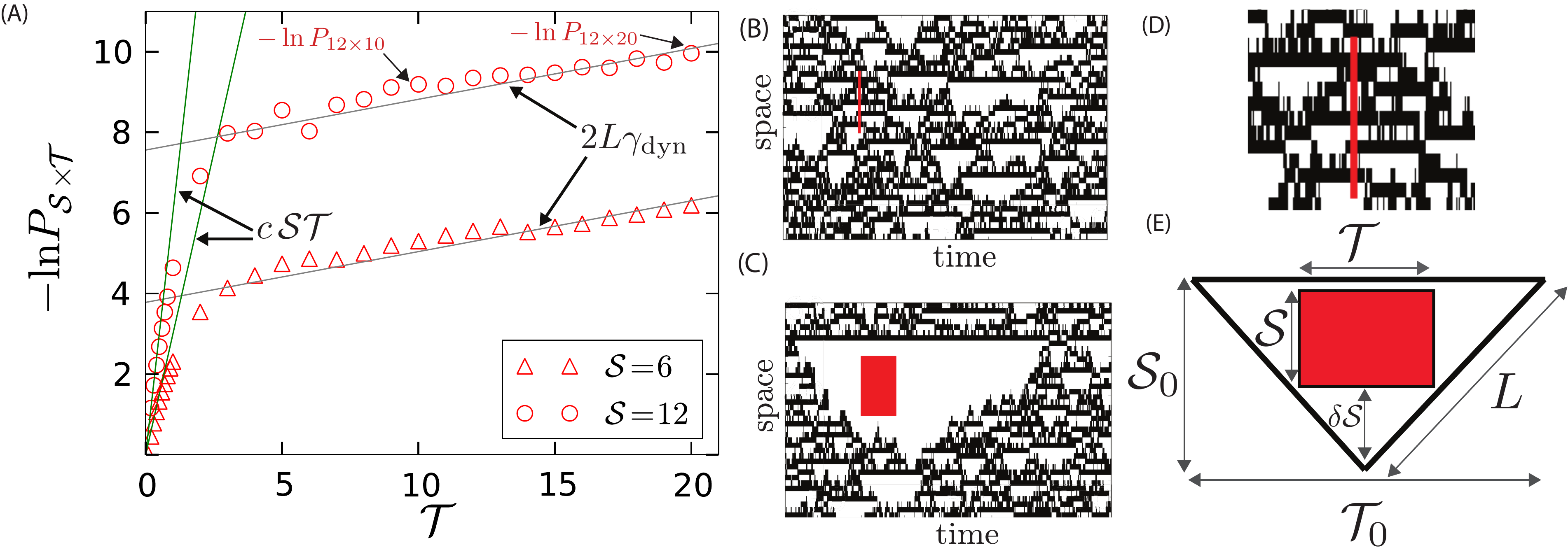}
\caption{(A) Dynamical free energy of solvation, $- \ln P_{\mathcal{S}\times\mathcal{T}}$ 
as a function of solute time size $\mathcal{T}$, for two spatial sizes $\mathcal{S}=12$ and $6$. The free energy shows two regimes of ``solvation'' for which the predicted scalings are plotted as lines. (The free energy values for solutes with $\mathcal{T}=10$ and $20$ are highlighted, cf.\ Fig.~\ref{fig:assembly}.) (B) For a smaller solute, there is no inactive domain created in the vicinity of the solute ($\mathcal{T}=1)$. (C) A pre-transition layer is observed for a large solute ($\mathcal{T}=12)$.  (D) Zoom in of the solute in (B). (E) Schematic for a large solute (red) and its surrounding inactive domain which is approximated as an inverted triangle with base $\mathcal{T}_0$ and height $\mathcal{S}_0$.
}
\label{fig:dfe}
\end{figure*}

\noindent
{\bf \em Dynamical free energy of solvation at small and large length scales.}
We can now utilize space--time interfaces to understand dynamical free energies of solvation of solutes in active trajectories, in analogy with the thermodynamic case, cf.\ Fig.~1. We define the dynamical free energy of solvation as the negative logarithm of the probability of finding trajectories with zero activity within the solute region, $-\ln P_{\mathcal{S}\times\mathcal{T}} = - \ln \left[ Z_{\mathrm{solute}=\mathcal{S}\times\mathcal{T}}(s,t_\mathrm{obs})/Z(s,t_\mathrm{obs}) \right]$.  The probabilities $P_{\mathcal{S}\times\mathcal{T}}$ are calculated using the path sampling procedure in Ref.~\cite{jack2006space}. 
As shown in Fig.~\ref{fig:dfe}, the dynamical free energy for solutes with different temporal sizes $\mathcal{T}$ shows two well-defined regimes of solvation with a crossover between them, as anticipated%, just as in the free energy of solvation of hard spheres in water~\cite{lum1999hydrophobicity,chandler2005interfaces,huang2001scaling}
, cf.\ Fig.~1. 
%
%In water, small fluctuations in density are sufficient to evacuate the molecules in the solute region yielding a small solute regime which scales as the volume of the solute. We can perform a similar analysis in our dynamics for small space--time solutes,

To derive the scaling for the small solute regime, cf.\ Fig.~4B,\,D, we consider the average number of excitations in the solute region given by $c\,\mathcal{S}$. As a consequence of the asymmetric kinetic constraint in this model, every excitation facilitates its right neighbor. Therefore, if the solute has $c\,\mathcal{S}$ excitations there are $c\,\mathcal{S}$ facilitated spins (i.e., spins that are unconstrained to flip) in the solute region, and at $T=\infty$ this is also the total rate of flipping spins within the solute region. 
If $t_\mathrm{wait}$ is the waiting time between flips, ${\rm Prob}(t_\mathrm{wait}\geqslant \mathcal{T}) = \exp(-c\,\mathcal{S} \mathcal{T})$ is the probability of observing a solute of time extent $\mathcal{T}$, given that $t_\mathrm{wait}$ is exponentially distributed with escape rate $c\,\mathcal{S}$. %The is the same as the probability of observing a solute of time extent $\mathcal{T}$. 
This results in a dynamical free energy for small solutes equal to $-\ln P_{\mathcal{S}\times\mathcal{T}}=c\, \mathcal{S} \mathcal{T}$ which scales as the area of the solute. This scaling is plotted in Fig.~\ref{fig:dfe}A and matches the numerically computed free energies for two different values of $\mathcal{S}$, showing that stochastic, uncorrelated fluctuations are sufficient to evacuate activity from within a small solute.

For the large solute regime 
%in water, the free energy of solvation scales as the surface area of the solute~\cite{lum1999hydrophobicity,chandler2005interfaces} because of the interfacial tension associated with the solute-induced vapor region. Analogously, in the case of active trajectories, 
we find that the dynamical free energy scales as the average length of the interface between the solute-induced inactive domain and the active bulk. The dynamical free energy of solvation can therefore be estimated given the length of the induced interface, and the proportionality constant can be interpreted as a dynamical interfacial tension~\cite{bodineau2012activity,bodineau2012finite,hedges2009dynamic,elmatad2010finite,nemoto2017finite}. 
The interfaces created in trajectories of the East model are triangular because of the asymmetry of the kinetic constraint, and we can estimate the length of the interface given a solute size from the sketch in Fig.~\ref{fig:dfe}E. The diagonal length $L$ is approximately $L=\sqrt{\mathcal{S}_0^2 + (\mathcal{T}_0/2)^2}$, where $\mathcal{S}_0 = \mathcal{S} + \delta S$, and $\delta S \approx \nu \mathcal{T}/2$ with $\nu$ being the interfacial velocity.
This interfacial velocity is the front velocity that can be calculated between the two bulk phases at coexistence in Fig.~\ref{fig:interface}A. The time $\mathcal{T}_0$ can be approximated to be $\mathcal{T}_0/2 \approx \mathcal{S}_0/\nu$. The total interfacial length is then given by $2L = 2 \, \sqrt{1+1/\nu^2} \, (\mathcal{S} + \nu\mathcal{T}/2)$. 
If, as we expect, the dynamical free energy of solvation for large space--time solutes is dominated by the interfacial free energy, then $-\ln P_{\mathcal{S}\times\mathcal{T}} = 2L \, \gamma_\mathrm{dyn}$, where $\gamma_\mathrm{dyn}$ is the ``dynamical'' interfacial tension. This leads to a prediction for the free energy $-\ln P_{\mathcal{S}\times\mathcal{T}} = \gamma_\mathrm{dyn} (2\sqrt{1+1/\nu^2)}(\mathcal{S} + \nu\mathcal{T}/2)$ which is linear in $\mathcal{T}$, as confirmed in Fig.~\ref{fig:dfe}A. This linear dependence yields a value of $\gamma_\mathrm{dyn} = 0.117$ from the slope of the large length scale regime for $\mathcal{S}=12$ in Fig.~\ref{fig:dfe}A. Using the same value of $\gamma_\mathrm{dyn}$, we predict the large length scale solvation free energy for a different solute size $\mathcal{S}=6$, again confirmed in Fig.~\ref{fig:dfe}A. 

{The dynamical interfacial tension calculated here is anisotropic and can be decomposed into a spatial component $\gamma_\mathrm{dyn} \sqrt{1+1/\nu^2}$ and a temporal component $\gamma_\mathrm{dyn} \nu (\sqrt{1+1/\nu^2})/2$. The latter is comparable to the interfacial tension proposed in Ref.~\cite{bodineau2012activity} and calculated in Ref.~\cite{bodineau2012finite}.}
\vspace{0.07in}

\begin{figure}[t] 
\includegraphics[width=1.0\columnwidth]{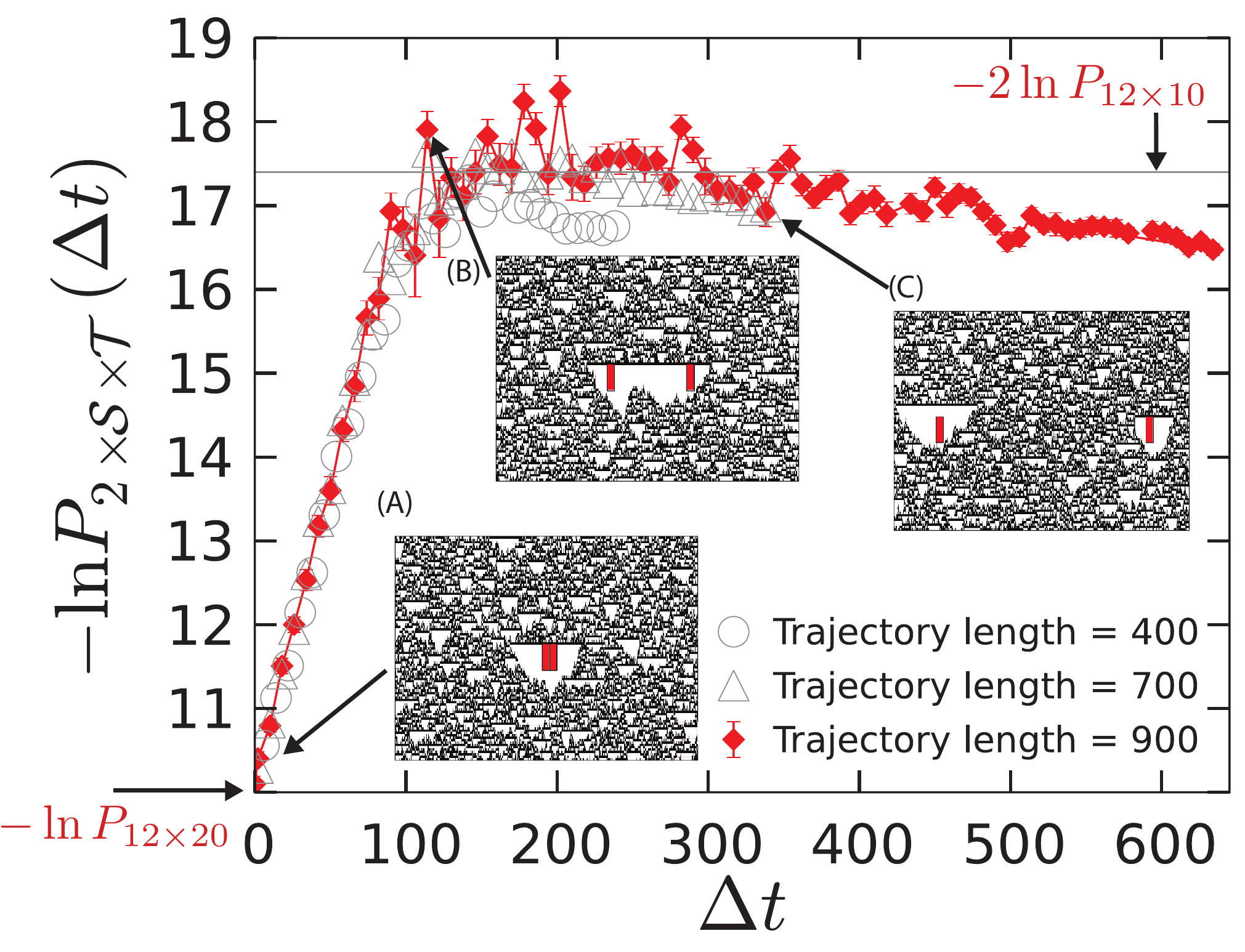}
\caption{Potential of mean force $- \ln P_{2\times\mathcal{S}\times\mathcal{T}}(\Delta t)$ between two large solutes as a function of temporal distance $\Delta t$ for three different total trajectory lengths. Here, two solutes with $\mathcal{S}=12$ and $\mathcal{T}=10$ effectively form a larger solute with $\mathcal{T}=20$ when they are together. Representative trajectory are shown when the solutes are close together (A), at an intermediate distance (B), and further apart (C). It can be seen that the two interfaces merge giving rise to a single interface. The effective free energy difference between the assembled and unassembled states agrees with the theoretical predictions shown as the gray line, as the total trajectory length increases. The values of $-\ln P_{12\times10}$ and $-\ln P_{12\times20}$ are taken from Fig.~\ref{fig:dfe}A.
} 
\label{fig:assembly}.
\end{figure}

\noindent
{\bf \em Dynamical ``solute assembly''.}
%An important consequence of the hydrophobic effect is {\em hydrophobic collapse}: large, hard solutes in water assemble due to the free energy gain of reducing the overall vapor--liquid interfacial area~\cite{chandler2005interfaces,willard2008role}.  
We now consider the dynamical free energy of `assembly' between two space--time solutes. If $P_{2\times\mathcal{S}\times\mathcal{T}}(\Delta t)$ is the probability of observing two space--time solutes of size $\mathcal{S}\times\mathcal{T}$, located at the same position in space, but separated in time by $\Delta t$, then the dynamical free energy for their solvation is $- \ln P_{2\times\mathcal{S}\times\mathcal{T}}(\Delta t)$.  We can write this free energy as a shift from that of two independent (i.e., distant) solutes, 
$- \ln P_{2\times\mathcal{S}\times\mathcal{T}}(\Delta t) = - 2 \ln P_{\mathcal{S}\times\mathcal{T}} + \gamma_\mathrm{dyn} \ell(\Delta t)$, where $\ell(\Delta t)$ represents the difference between the active--inactive interfacial lengths in the two cases. From this we expect that the most probable state is one where the solutes are together as $\ell$ will be the smallest at $\Delta t =0$. Figure~\ref{fig:assembly} shows the dynamical free energy between the solutes as a function of their time separation $\Delta t$ and confirms our prediction: trajectories with solutes ``assembled'' indeed have a higher probability compared to trajectories with the solutes further apart.  In fact, the free energy difference between the assembled ($\Delta t = 0$) and disassembled ($\Delta t \gg \mathcal{T}$) states tends to $- 2 \ln P_{\mathcal{S}\times\mathcal{T}} +  \ln P_{\mathcal{S}\times 2\mathcal{T}}$, as expected.

\vspace{0.07in}

\noindent
{\bf \em Conclusion.}
We have demonstrated dynamical pre-transition phenomena that are the trajectory space equivalent of hydrophobicity in liquid water and of similar physics in systems close to a first-order thermodynamic transition. We illustrated our results with the East model at infinite temperature, but we expect similar behavior in systems with slow, collective relaxation. These include other kinetically constrained models~\cite{garrahan2003coarse,fredrickson1984kinetic,elmatad2013space} and atomistic liquids in their supercooled regime~\cite{keys2011}, as well as a wider range of classical and open quantum problems displaying intermittent dynamics that have dynamical first-order transitions. {These include exclusion processes and driven diffusive systems~\cite{lazarescu2017generic,baek2017dynamical,shpielberg2017numerical,tizon2017order,derrida1998exactly}, protein dynamics~\cite{weber2013emergence,weber2014dynamical}, quantum systems~\cite{lesanovsky2013characterization,heyl2017dynamical}, as well as active matter~\cite{cagnetta2017large}.} Beyond the interesting conceptual analogies between dynamic and thermodynamic hydrophobic-like physics, the results presented here should have clear observable consequences. For example, the space--time solute ``assembly'', cf.\ Fig.~5, has implications for the relative frequency of occurrence of dynamically inactive regions in systems with heterogeneous dynamics, such as supercooled liquids. {Specifically, for a large enough size of inactive region in space--time in a supercooled liquid, it is more probable to observe two such regions in close succession rather than separated in space--time. The relative probability of these two types of trajectories can be used to determine a dynamical interfacial tension.} Such predictions could be tested in experiments~\cite{gokhale2014} and in atomistic simulations~\cite{keys2011} of supercooled liquids giving further validation to the idea that slow heterogeneous dynamics occurs due to proximity to a dynamical phase transition~\cite{chandler2010dynamics}. 

\bigskip

\begin{acknowledgments}
We are indebted to David Chandler for extended and lively discussions on the hydrophobic effect as well as glasses. We would like to thank David Limmer for useful discussions. S.K. acknowledges funding from the National Science Foundation (award number 1507642). K.K.M acknowledges funding from the Department of Energy (contract DE-AC02-05CH11231, FWP No. CHPHYS02). JPG was supported by EPSRC Grant No.\ EP/M014266/1.
\end{acknowledgments}

%\bibliography{PreTransitionEast}

\begin{thebibliography}{37}%
\makeatletter
\providecommand \@ifxundefined [1]{%
 \@ifx{#1\undefined}
}%
\providecommand \@ifnum [1]{%
 \ifnum #1\expandafter \@firstoftwo
 \else \expandafter \@secondoftwo
 \fi
}%
\providecommand \@ifx [1]{%
 \ifx #1\expandafter \@firstoftwo
 \else \expandafter \@secondoftwo
 \fi
}%
\providecommand \natexlab [1]{#1}%
\providecommand \enquote  [1]{``#1''}%
\providecommand \bibnamefont  [1]{#1}%
\providecommand \bibfnamefont [1]{#1}%
\providecommand \citenamefont [1]{#1}%
\providecommand \href@noop [0]{\@secondoftwo}%
\providecommand \href [0]{\begingroup \@sanitize@url \@href}%
\providecommand \@href[1]{\@@startlink{#1}\@@href}%
\providecommand \@@href[1]{\endgroup#1\@@endlink}%
\providecommand \@sanitize@url [0]{\catcode `\\12\catcode `\$12\catcode
  `\&12\catcode `\#12\catcode `\^12\catcode `\_12\catcode `\%12\relax}%
\providecommand \@@startlink[1]{}%
\providecommand \@@endlink[0]{}%
\providecommand \url  [0]{\begingroup\@sanitize@url \@url }%
\providecommand \@url [1]{\endgroup\@href {#1}{\urlprefix }}%
\providecommand \urlprefix  [0]{URL }%
\providecommand \Eprint [0]{\href }%
\providecommand \doibase [0]{http://dx.doi.org/}%
\providecommand \selectlanguage [0]{\@gobble}%
\providecommand \bibinfo  [0]{\@secondoftwo}%
\providecommand \bibfield  [0]{\@secondoftwo}%
\providecommand \translation [1]{[#1]}%
\providecommand \BibitemOpen [0]{}%
\providecommand \bibitemStop [0]{}%
\providecommand \bibitemNoStop [0]{.\EOS\space}%
\providecommand \EOS [0]{\spacefactor3000\relax}%
\providecommand \BibitemShut  [1]{\csname bibitem#1\endcsname}%
\let\auto@bib@innerbib\@empty
%</preamble>
\bibitem [{\citenamefont {Lipowsky}(1982)}]{lipowsky1982critical}%
  \BibitemOpen
  \bibfield  {author} {\bibinfo {author} {\bibfnamefont {R.}~\bibnamefont
  {Lipowsky}},\ }\href@noop {} {\bibfield  {journal} {\bibinfo  {journal}
  {Physical Review Letters}\ }\textbf {\bibinfo {volume} {49}},\ \bibinfo
  {pages} {1575} (\bibinfo {year} {1982})}\BibitemShut {NoStop}%
\bibitem [{\citenamefont {Lipowsky}(1984)}]{lipowsky1984surface}%
  \BibitemOpen
  \bibfield  {author} {\bibinfo {author} {\bibfnamefont {R.}~\bibnamefont
  {Lipowsky}},\ }\href@noop {} {\bibfield  {journal} {\bibinfo  {journal}
  {Journal of Applied Physics}\ }\textbf {\bibinfo {volume} {55}},\ \bibinfo
  {pages} {2485} (\bibinfo {year} {1984})}\BibitemShut {NoStop}%
\bibitem [{\citenamefont {Lum}\ \emph {et~al.}(1999)\citenamefont {Lum},
  \citenamefont {Chandler},\ and\ \citenamefont
  {Weeks}}]{lum1999hydrophobicity}%
  \BibitemOpen
  \bibfield  {author} {\bibinfo {author} {\bibfnamefont {K.}~\bibnamefont
  {Lum}}, \bibinfo {author} {\bibfnamefont {D.}~\bibnamefont {Chandler}}, \
  and\ \bibinfo {author} {\bibfnamefont {J.~D.}\ \bibnamefont {Weeks}},\
  }\href@noop {} {\bibfield  {journal} {\bibinfo  {journal} {The Journal of
  Physical Chemistry B}\ }\textbf {\bibinfo {volume} {103}},\ \bibinfo {pages}
  {4570} (\bibinfo {year} {1999})}\BibitemShut {NoStop}%
\bibitem [{\citenamefont {Chandler}(2005)}]{chandler2005interfaces}%
  \BibitemOpen
  \bibfield  {author} {\bibinfo {author} {\bibfnamefont {D.}~\bibnamefont
  {Chandler}},\ }\href@noop {} {\bibfield  {journal} {\bibinfo  {journal}
  {Nature}\ }\textbf {\bibinfo {volume} {437}},\ \bibinfo {pages} {640}
  (\bibinfo {year} {2005})}\BibitemShut {NoStop}%
\bibitem [{\citenamefont {Katira}\ \emph {et~al.}(2016)\citenamefont {Katira},
  \citenamefont {Mandadapu}, \citenamefont {Vaikuntanathan}, \citenamefont
  {Smit},\ and\ \citenamefont {Chandler}}]{katira2016pre}%
  \BibitemOpen
  \bibfield  {author} {\bibinfo {author} {\bibfnamefont {S.}~\bibnamefont
  {Katira}}, \bibinfo {author} {\bibfnamefont {K.~K.}\ \bibnamefont
  {Mandadapu}}, \bibinfo {author} {\bibfnamefont {S.}~\bibnamefont
  {Vaikuntanathan}}, \bibinfo {author} {\bibfnamefont {B.}~\bibnamefont
  {Smit}}, \ and\ \bibinfo {author} {\bibfnamefont {D.}~\bibnamefont
  {Chandler}},\ }\href@noop {} {\bibfield  {journal} {\bibinfo  {journal}
  {Elife}\ }\textbf {\bibinfo {volume} {5}},\ \bibinfo {pages} {e13150}
  (\bibinfo {year} {2016})}\BibitemShut {NoStop}%
\bibitem [{\citenamefont {Huang}\ and\ \citenamefont
  {Chandler}(2000)}]{huang2000cavity}%
  \BibitemOpen
  \bibfield  {author} {\bibinfo {author} {\bibfnamefont {D.~M.}\ \bibnamefont
  {Huang}}\ and\ \bibinfo {author} {\bibfnamefont {D.}~\bibnamefont
  {Chandler}},\ }\href@noop {} {\bibfield  {journal} {\bibinfo  {journal}
  {Physical Review E}\ }\textbf {\bibinfo {volume} {61}},\ \bibinfo {pages}
  {1501} (\bibinfo {year} {2000})}\BibitemShut {NoStop}%
\bibitem [{\citenamefont {Huang}\ \emph {et~al.}(2001)\citenamefont {Huang},
  \citenamefont {Geissler},\ and\ \citenamefont {Chandler}}]{huang2001scaling}%
  \BibitemOpen
  \bibfield  {author} {\bibinfo {author} {\bibfnamefont {D.~M.}\ \bibnamefont
  {Huang}}, \bibinfo {author} {\bibfnamefont {P.~L.}\ \bibnamefont {Geissler}},
  \ and\ \bibinfo {author} {\bibfnamefont {D.}~\bibnamefont {Chandler}},\
  }\href@noop {} {\bibfield  {journal} {\bibinfo  {journal} {The Journal of
  Physical Chemistry B}\ }\textbf {\bibinfo {volume} {105}},\ \bibinfo {pages}
  {6704} (\bibinfo {year} {2001})}\BibitemShut {NoStop}%
\bibitem [{\citenamefont {Mittal}\ and\ \citenamefont
  {Hummer}(2008)}]{mittal2008static}%
  \BibitemOpen
  \bibfield  {author} {\bibinfo {author} {\bibfnamefont {J.}~\bibnamefont
  {Mittal}}\ and\ \bibinfo {author} {\bibfnamefont {G.}~\bibnamefont
  {Hummer}},\ }\href@noop {} {\bibfield  {journal} {\bibinfo  {journal}
  {Proceedings of the National Academy of Sciences}\ }\textbf {\bibinfo
  {volume} {105}},\ \bibinfo {pages} {20130} (\bibinfo {year}
  {2008})}\BibitemShut {NoStop}%
\bibitem [{\citenamefont {Willard}\ and\ \citenamefont
  {Chandler}(2008)}]{willard2008role}%
  \BibitemOpen
  \bibfield  {author} {\bibinfo {author} {\bibfnamefont {A.~P.}\ \bibnamefont
  {Willard}}\ and\ \bibinfo {author} {\bibfnamefont {D.}~\bibnamefont
  {Chandler}},\ }\href@noop {} {\bibfield  {journal} {\bibinfo  {journal} {The
  Journal of Physical Chemistry B}\ }\textbf {\bibinfo {volume} {112}},\
  \bibinfo {pages} {6187} (\bibinfo {year} {2008})}\BibitemShut {NoStop}%
\bibitem [{\citenamefont {Elmatad}\ and\ \citenamefont
  {Jack}(2013)}]{elmatad2013space}%
  \BibitemOpen
  \bibfield  {author} {\bibinfo {author} {\bibfnamefont {Y.~S.}\ \bibnamefont
  {Elmatad}}\ and\ \bibinfo {author} {\bibfnamefont {R.~L.}\ \bibnamefont
  {Jack}},\ }\href@noop {} {\bibfield  {journal} {\bibinfo  {journal} {The
  Journal of Chemical Physics}\ }\textbf {\bibinfo {volume} {138}},\ \bibinfo
  {pages} {12A531} (\bibinfo {year} {2013})}\BibitemShut {NoStop}%
  \bibitem [{\citenamefont {Lazarescu}(2017)}]{lazarescu2017generic}%
  \BibitemOpen
  \bibfield  {author} {\bibinfo {author} {\bibfnamefont {A.}~\bibnamefont
  {Lazarescu}},\ }\href@noop {} {\bibfield  {journal} {\bibinfo  {journal}
  {Journal of Physics A: Mathematical and Theoretical}\ }\textbf {\bibinfo
  {volume} {50}},\ \bibinfo {pages} {254004} (\bibinfo {year}
  {2017})}\BibitemShut {NoStop}%
\bibitem [{\citenamefont {Baek}\ \emph {et~al.}(2017)\citenamefont {Baek},
  \citenamefont {Kafri},\ and\ \citenamefont {Lecomte}}]{baek2017dynamical}%
  \BibitemOpen
  \bibfield  {author} {\bibinfo {author} {\bibfnamefont {Y.}~\bibnamefont
  {Baek}}, \bibinfo {author} {\bibfnamefont {Y.}~\bibnamefont {Kafri}}, \ and\
  \bibinfo {author} {\bibfnamefont {V.}~\bibnamefont {Lecomte}},\ }\href@noop
  {} {\bibfield  {journal} {\bibinfo  {journal} {Physical review letters}\
  }\textbf {\bibinfo {volume} {118}},\ \bibinfo {pages} {030604} (\bibinfo
  {year} {2017})}\BibitemShut {NoStop}%
\bibitem [{\citenamefont {Shpielberg}\ \emph {et~al.}(2017)\citenamefont
  {Shpielberg}, \citenamefont {Don},\ and\ \citenamefont
  {Akkermans}}]{shpielberg2017numerical}%
  \BibitemOpen
  \bibfield  {author} {\bibinfo {author} {\bibfnamefont {O.}~\bibnamefont
  {Shpielberg}}, \bibinfo {author} {\bibfnamefont {Y.}~\bibnamefont {Don}}, \
  and\ \bibinfo {author} {\bibfnamefont {E.}~\bibnamefont {Akkermans}},\
  }\href@noop {} {\bibfield  {journal} {\bibinfo  {journal} {Physical Review
  E}\ }\textbf {\bibinfo {volume} {95}},\ \bibinfo {pages} {032137} (\bibinfo
  {year} {2017})}\BibitemShut {NoStop}%
\bibitem [{\citenamefont {Bertini}\ \emph {et~al.}(2015)\citenamefont
  {Bertini}, \citenamefont {De~Sole}, \citenamefont {Gabrielli}, \citenamefont
  {Jona-Lasinio},\ and\ \citenamefont {Landim}}]{bertini2015macroscopic}%
  \BibitemOpen
  \bibfield  {author} {\bibinfo {author} {\bibfnamefont {L.}~\bibnamefont
  {Bertini}}, \bibinfo {author} {\bibfnamefont {A.}~\bibnamefont {De~Sole}},
  \bibinfo {author} {\bibfnamefont {D.}~\bibnamefont {Gabrielli}}, \bibinfo
  {author} {\bibfnamefont {G.}~\bibnamefont {Jona-Lasinio}}, \ and\ \bibinfo
  {author} {\bibfnamefont {C.}~\bibnamefont {Landim}},\ }\href@noop {}
  {\bibfield  {journal} {\bibinfo  {journal} {Reviews of Modern Physics}\
  }\textbf {\bibinfo {volume} {87}},\ \bibinfo {pages} {593} (\bibinfo {year}
  {2015})}\BibitemShut {NoStop}%
\bibitem [{\citenamefont {Bodineau}\ and\ \citenamefont
  {Derrida}(2004)}]{bodineau2004current}%
  \BibitemOpen
  \bibfield  {author} {\bibinfo {author} {\bibfnamefont {T.}~\bibnamefont
  {Bodineau}}\ and\ \bibinfo {author} {\bibfnamefont {B.}~\bibnamefont
  {Derrida}},\ }\href@noop {} {\bibfield  {journal} {\bibinfo  {journal}
  {Physical Review Letters}\ }\textbf {\bibinfo {volume} {92}},\ \bibinfo
  {pages} {180601} (\bibinfo {year} {2004})}\BibitemShut {NoStop}%
\bibitem [{\citenamefont {Bodineau}\ and\ \citenamefont
  {Derrida}(2005)}]{bodineau2005distribution}%
  \BibitemOpen
  \bibfield  {author} {\bibinfo {author} {\bibfnamefont {T.}~\bibnamefont
  {Bodineau}}\ and\ \bibinfo {author} {\bibfnamefont {B.}~\bibnamefont
  {Derrida}},\ }\href@noop {} {\bibfield  {journal} {\bibinfo  {journal}
  {Physical Review E}\ }\textbf {\bibinfo {volume} {72}},\ \bibinfo {pages}
  {066110} (\bibinfo {year} {2005})}\BibitemShut {NoStop}%
\bibitem [{\citenamefont {Hurtado}\ and\ \citenamefont
  {Garrido}(2011)}]{hurtado2011spontaneous}%
  \BibitemOpen
  \bibfield  {author} {\bibinfo {author} {\bibfnamefont {P.~I.}\ \bibnamefont
  {Hurtado}}\ and\ \bibinfo {author} {\bibfnamefont {P.~L.}\ \bibnamefont
  {Garrido}},\ }\href@noop {} {\bibfield  {journal} {\bibinfo  {journal}
  {Physical Review Letters}\ }\textbf {\bibinfo {volume} {107}},\ \bibinfo
  {pages} {180601} (\bibinfo {year} {2011})}\BibitemShut {NoStop}%
\bibitem [{\citenamefont {Lecomte}\ \emph {et~al.}(2007)\citenamefont
  {Lecomte}, \citenamefont {Appert-Rolland},\ and\ \citenamefont
  {Van~Wijland}}]{lecomte2007thermodynamic}%
  \BibitemOpen
  \bibfield  {author} {\bibinfo {author} {\bibfnamefont {V.}~\bibnamefont
  {Lecomte}}, \bibinfo {author} {\bibfnamefont {C.}~\bibnamefont
  {Appert-Rolland}}, \ and\ \bibinfo {author} {\bibfnamefont {F.}~\bibnamefont
  {Van~Wijland}},\ }\href@noop {} {\bibfield  {journal} {\bibinfo  {journal}
  {Journal of Statistical Physics}\ }\textbf {\bibinfo {volume} {127}},\
  \bibinfo {pages} {51} (\bibinfo {year} {2007})}\BibitemShut {NoStop}%
\bibitem [{\citenamefont {Garrahan}\ \emph {et~al.}(2007)\citenamefont
  {Garrahan}, \citenamefont {Jack}, \citenamefont {Lecomte}, \citenamefont
  {Pitard}, \citenamefont {van Duijvendijk},\ and\ \citenamefont {van
  Wijland}}]{garrahan2007dynamical}%
  \BibitemOpen
  \bibfield  {author} {\bibinfo {author} {\bibfnamefont {J.~P.}\ \bibnamefont
  {Garrahan}}, \bibinfo {author} {\bibfnamefont {R.~L.}\ \bibnamefont {Jack}},
  \bibinfo {author} {\bibfnamefont {V.}~\bibnamefont {Lecomte}}, \bibinfo
  {author} {\bibfnamefont {E.}~\bibnamefont {Pitard}}, \bibinfo {author}
  {\bibfnamefont {K.}~\bibnamefont {van Duijvendijk}}, \ and\ \bibinfo {author}
  {\bibfnamefont {F.}~\bibnamefont {van Wijland}},\ }\href@noop {} {\bibfield
  {journal} {\bibinfo  {journal} {Physical Review Letters}\ }\textbf {\bibinfo
  {volume} {98}},\ \bibinfo {pages} {195702} (\bibinfo {year}
  {2007})}\BibitemShut {NoStop}%
\bibitem [{\citenamefont {Hedges}\ \emph {et~al.}(2009)\citenamefont {Hedges},
  \citenamefont {Jack}, \citenamefont {Garrahan},\ and\ \citenamefont
  {Chandler}}]{hedges2009dynamic}%
  \BibitemOpen
  \bibfield  {author} {\bibinfo {author} {\bibfnamefont {L.~O.}\ \bibnamefont
  {Hedges}}, \bibinfo {author} {\bibfnamefont {R.~L.}\ \bibnamefont {Jack}},
  \bibinfo {author} {\bibfnamefont {J.~P.}\ \bibnamefont {Garrahan}}, \ and\
  \bibinfo {author} {\bibfnamefont {D.}~\bibnamefont {Chandler}},\ }\href@noop
  {} {\bibfield  {journal} {\bibinfo  {journal} {Science}\ }\textbf {\bibinfo
  {volume} {323}},\ \bibinfo {pages} {1309} (\bibinfo {year}
  {2009})}\BibitemShut {NoStop}%
\bibitem [{\citenamefont {Limmer}\ and\ \citenamefont
  {Chandler}(2014)}]{limmer2014theory}%
  \BibitemOpen
  \bibfield  {author} {\bibinfo {author} {\bibfnamefont {D.~T.}\ \bibnamefont
  {Limmer}}\ and\ \bibinfo {author} {\bibfnamefont {D.}~\bibnamefont
  {Chandler}},\ }\href@noop {} {\bibfield  {journal} {\bibinfo  {journal}
  {Proceedings of the National Academy of Sciences}\ }\textbf {\bibinfo
  {volume} {111}},\ \bibinfo {pages} {9413} (\bibinfo {year}
  {2014})}\BibitemShut {NoStop}%
\bibitem [{\citenamefont {Vaikuntanathan}\ \emph {et~al.}(2014)\citenamefont
  {Vaikuntanathan}, \citenamefont {Gingrich},\ and\ \citenamefont
  {Geissler}}]{vaikuntanathan2014dynamic}%
  \BibitemOpen
  \bibfield  {author} {\bibinfo {author} {\bibfnamefont {S.}~\bibnamefont
  {Vaikuntanathan}}, \bibinfo {author} {\bibfnamefont {T.~R.}\ \bibnamefont
  {Gingrich}}, \ and\ \bibinfo {author} {\bibfnamefont {P.~L.}\ \bibnamefont
  {Geissler}},\ }\href@noop {} {\bibfield  {journal} {\bibinfo  {journal}
  {Physical Review E}\ }\textbf {\bibinfo {volume} {89}},\ \bibinfo {pages}
  {062108} (\bibinfo {year} {2014})}\BibitemShut {NoStop}%
\bibitem [{\citenamefont {Tiz{\'o}n-Escamilla}\ \emph
  {et~al.}(2017)\citenamefont {Tiz{\'o}n-Escamilla}, \citenamefont
  {P{\'e}rez-Espigares}, \citenamefont {Garrido},\ and\ \citenamefont
  {Hurtado}}]{tizon2017order}%
  \BibitemOpen
  \bibfield  {author} {\bibinfo {author} {\bibfnamefont {N.}~\bibnamefont
  {Tiz{\'o}n-Escamilla}}, \bibinfo {author} {\bibfnamefont {C.}~\bibnamefont
  {P{\'e}rez-Espigares}}, \bibinfo {author} {\bibfnamefont {P.~L.}\
  \bibnamefont {Garrido}}, \ and\ \bibinfo {author} {\bibfnamefont {P.~I.}\
  \bibnamefont {Hurtado}},\ }\href@noop {} {\bibfield  {journal} {\bibinfo
  {journal} {Physical Review Letters}\ }\textbf {\bibinfo {volume} {119}},\
  \bibinfo {pages} {090602} (\bibinfo {year} {2017})}\BibitemShut {NoStop}%
\bibitem [{\citenamefont {Garrahan}\ and\ \citenamefont
  {Lesanovsky}(2010)}]{garrahan2010thermodynamics}%
  \BibitemOpen
  \bibfield  {author} {\bibinfo {author} {\bibfnamefont {J.~P.}\ \bibnamefont
  {Garrahan}}\ and\ \bibinfo {author} {\bibfnamefont {I.}~\bibnamefont
  {Lesanovsky}},\ }\href@noop {} {\bibfield  {journal} {\bibinfo  {journal}
  {Physical Review Letters}\ }\textbf {\bibinfo {volume} {104}},\ \bibinfo
  {pages} {160601} (\bibinfo {year} {2010})}\BibitemShut {NoStop}%
\bibitem [{\citenamefont {Lesanovsky}\ \emph {et~al.}(2013)\citenamefont
  {Lesanovsky}, \citenamefont {van Horssen}, \citenamefont {Gu{\c{t}}{\u{a}}},\
  and\ \citenamefont {Garrahan}}]{lesanovsky2013characterization}%
  \BibitemOpen
  \bibfield  {author} {\bibinfo {author} {\bibfnamefont {I.}~\bibnamefont
  {Lesanovsky}}, \bibinfo {author} {\bibfnamefont {M.}~\bibnamefont {van
  Horssen}}, \bibinfo {author} {\bibfnamefont {M.}~\bibnamefont
  {Gu{\c{t}}{\u{a}}}}, \ and\ \bibinfo {author} {\bibfnamefont {J.~P.}\
  \bibnamefont {Garrahan}},\ }\href@noop {} {\bibfield  {journal} {\bibinfo
  {journal} {Physical Review Letters}\ }\textbf {\bibinfo {volume} {110}},\
  \bibinfo {pages} {150401} (\bibinfo {year} {2013})}\BibitemShut {NoStop}%
\bibitem [{\citenamefont {Ates}\ \emph {et~al.}(2012)\citenamefont {Ates},
  \citenamefont {Olmos}, \citenamefont {Garrahan},\ and\ \citenamefont
  {Lesanovsky}}]{ates2012dynamical}%
  \BibitemOpen
  \bibfield  {author} {\bibinfo {author} {\bibfnamefont {C.}~\bibnamefont
  {Ates}}, \bibinfo {author} {\bibfnamefont {B.}~\bibnamefont {Olmos}},
  \bibinfo {author} {\bibfnamefont {J.~P.}\ \bibnamefont {Garrahan}}, \ and\
  \bibinfo {author} {\bibfnamefont {I.}~\bibnamefont {Lesanovsky}},\
  }\href@noop {} {\bibfield  {journal} {\bibinfo  {journal} {Physical Review
  A}\ }\textbf {\bibinfo {volume} {85}},\ \bibinfo {pages} {043620} (\bibinfo
  {year} {2012})}\BibitemShut {NoStop}%
\bibitem [{\citenamefont {Garrahan}\ \emph {et~al.}(2009)\citenamefont
  {Garrahan}, \citenamefont {Jack}, \citenamefont {Lecomte}, \citenamefont
  {Pitard}, \citenamefont {van Duijvendijk},\ and\ \citenamefont {van
  Wijland}}]{garrahan2009first}%
  \BibitemOpen
  \bibfield  {author} {\bibinfo {author} {\bibfnamefont {J.~P.}\ \bibnamefont
  {Garrahan}}, \bibinfo {author} {\bibfnamefont {R.~L.}\ \bibnamefont {Jack}},
  \bibinfo {author} {\bibfnamefont {V.}~\bibnamefont {Lecomte}}, \bibinfo
  {author} {\bibfnamefont {E.}~\bibnamefont {Pitard}}, \bibinfo {author}
  {\bibfnamefont {K.}~\bibnamefont {van Duijvendijk}}, \ and\ \bibinfo {author}
  {\bibfnamefont {F.}~\bibnamefont {van Wijland}},\ }\href@noop {} {\bibfield
  {journal} {\bibinfo  {journal} {Journal of Physics A: Mathematical and
  Theoretical}\ }\textbf {\bibinfo {volume} {42}},\ \bibinfo {pages} {075007}
  (\bibinfo {year} {2009})}\BibitemShut {NoStop}%
\bibitem [{\citenamefont {J{\"a}ckle}\ and\ \citenamefont
  {Eisinger}(1991)}]{jackle1991hierarchically}%
  \BibitemOpen
  \bibfield  {author} {\bibinfo {author} {\bibfnamefont {J.}~\bibnamefont
  {J{\"a}ckle}}\ and\ \bibinfo {author} {\bibfnamefont {S.}~\bibnamefont
  {Eisinger}},\ }\href@noop {} {\bibfield  {journal} {\bibinfo  {journal}
  {Zeitschrift f{\"u}r Physik B Condensed Matter}\ }\textbf {\bibinfo {volume}
  {84}},\ \bibinfo {pages} {115} (\bibinfo {year} {1991})}\BibitemShut
  {NoStop}%
\bibitem [{Note1()}]{Note1}%
  \BibitemOpen
  \bibinfo {note} {The energy function for the East model is defined as
  $H=\DOTSB \sum@ \slimits@ _i n_i$. The concentration of excited spins is
  given by $c=$$ \langle n_i \rangle $$=\protect \qopname \relax
  o{exp}(-1/T)/(1+\protect \qopname \relax
  o{exp}(-1/T))$, $c$ is 1/2 at $T=\infty $}\BibitemShut {NoStop}%
\bibitem [{\citenamefont {Jack}\ \emph {et~al.}(2006)\citenamefont {Jack},
  \citenamefont {Garrahan},\ and\ \citenamefont {Chandler}}]{jack2006space}%
  \BibitemOpen
  \bibfield  {author} {\bibinfo {author} {\bibfnamefont {R.~L.}\ \bibnamefont
  {Jack}}, \bibinfo {author} {\bibfnamefont {J.~P.}\ \bibnamefont {Garrahan}},
  \ and\ \bibinfo {author} {\bibfnamefont {D.}~\bibnamefont {Chandler}},\
  }\href@noop {} {\bibfield  {journal} {\bibinfo  {journal} {The Journal of
  Chemical Physics}\ }\textbf {\bibinfo {volume} {125}},\ \bibinfo {pages}
  {184509} (\bibinfo {year} {2006})}\BibitemShut {NoStop}%
  \bibitem [{\citenamefont {Rowlinson}\ and\ \citenamefont
  {Widom}(2013)}]{rowlinson2013molecular}%
  \BibitemOpen
  \bibfield  {author} {\bibinfo {author} {\bibfnamefont {J.~S.}\ \bibnamefont
  {Rowlinson}}\ and\ \bibinfo {author} {\bibfnamefont {B.}~\bibnamefont
  {Widom}},\ }\href@noop {} {\emph {\bibinfo {title} {Molecular theory of
  capillarity}}}\ (\bibinfo  {publisher} {Courier Corporation},\ \bibinfo
  {year} {2013})\BibitemShut {NoStop}% 
\bibitem [{\citenamefont {Ganguly}\ \emph {et~al.}(2015)\citenamefont
  {Ganguly}, \citenamefont {Lubetzky},\ and\ \citenamefont
  {Martinelli}}]{ganguly2015cutoff}%
  \BibitemOpen
  \bibfield  {author} {\bibinfo {author} {\bibfnamefont {S.}~\bibnamefont
  {Ganguly}}, \bibinfo {author} {\bibfnamefont {E.}~\bibnamefont {Lubetzky}}, \
  and\ \bibinfo {author} {\bibfnamefont {F.}~\bibnamefont {Martinelli}},\
  }\href@noop {} {\bibfield  {journal} {\bibinfo  {journal} {Communications in
  Mathematical Physics}\ }\textbf {\bibinfo {volume} {335}},\ \bibinfo {pages}
  {1287} (\bibinfo {year} {2015})}\BibitemShut {NoStop}%
\bibitem [{\citenamefont {Bodineau}\ and\ \citenamefont
  {Toninelli}(2012)}]{bodineau2012activity}%
  \BibitemOpen
  \bibfield  {author} {\bibinfo {author} {\bibfnamefont {T.}~\bibnamefont
  {Bodineau}}\ and\ \bibinfo {author} {\bibfnamefont {C.}~\bibnamefont
  {Toninelli}},\ }\href@noop {} {\bibfield  {journal} {\bibinfo  {journal}
  {Communications in Mathematical Physics}\ }\textbf {\bibinfo {volume}
  {311}},\ \bibinfo {pages} {357} (\bibinfo {year} {2012})}\BibitemShut
  {NoStop}%
\bibitem [{\citenamefont {Bodineau}\ \emph {et~al.}(2012)\citenamefont
  {Bodineau}, \citenamefont {Lecomte},\ and\ \citenamefont
  {Toninelli}}]{bodineau2012finite}%
  \BibitemOpen
  \bibfield  {author} {\bibinfo {author} {\bibfnamefont {T.}~\bibnamefont
  {Bodineau}}, \bibinfo {author} {\bibfnamefont {V.}~\bibnamefont {Lecomte}}, \
  and\ \bibinfo {author} {\bibfnamefont {C.}~\bibnamefont {Toninelli}},\
  }\href@noop {} {\bibfield  {journal} {\bibinfo  {journal} {Journal of
  Statistical Physics}\ }\textbf {\bibinfo {volume} {147}},\ \bibinfo {pages}
  {1} (\bibinfo {year} {2012})}\BibitemShut {NoStop}%
\bibitem [{\citenamefont {Elmatad}\ \emph {et~al.}(2010)\citenamefont
  {Elmatad}, \citenamefont {Jack}, \citenamefont {Chandler},\ and\
  \citenamefont {Garrahan}}]{elmatad2010finite}%
  \BibitemOpen
  \bibfield  {author} {\bibinfo {author} {\bibfnamefont {Y.~S.}\ \bibnamefont
  {Elmatad}}, \bibinfo {author} {\bibfnamefont {R.~L.}\ \bibnamefont {Jack}},
  \bibinfo {author} {\bibfnamefont {D.}~\bibnamefont {Chandler}}, \ and\
  \bibinfo {author} {\bibfnamefont {J.~P.}\ \bibnamefont {Garrahan}},\
  }\href@noop {} {\bibfield  {journal} {\bibinfo  {journal} {Proceedings of the
  National Academy of Sciences}\ }\textbf {\bibinfo {volume} {107}},\ \bibinfo
  {pages} {12793} (\bibinfo {year} {2010})}\BibitemShut {NoStop}%
\bibitem [{\citenamefont {Nemoto}\ \emph {et~al.}(2017)\citenamefont {Nemoto},
  \citenamefont {Jack},\ and\ \citenamefont {Lecomte}}]{nemoto2017finite}%
  \BibitemOpen
  \bibfield  {author} {\bibinfo {author} {\bibfnamefont {T.}~\bibnamefont
  {Nemoto}}, \bibinfo {author} {\bibfnamefont {R.~L.}\ \bibnamefont {Jack}}, \
  and\ \bibinfo {author} {\bibfnamefont {V.}~\bibnamefont {Lecomte}},\
  }\href@noop {} {\bibfield  {journal} {\bibinfo  {journal} {Physical Review
  Letters}\ }\textbf {\bibinfo {volume} {118}},\ \bibinfo {pages} {115702}
  (\bibinfo {year} {2017})}\BibitemShut {NoStop}%
\bibitem [{\citenamefont {Garrahan}\ and\ \citenamefont
  {Chandler}(2003)}]{garrahan2003coarse}%
  \BibitemOpen
  \bibfield  {author} {\bibinfo {author} {\bibfnamefont {J.~P.}\ \bibnamefont
  {Garrahan}}\ and\ \bibinfo {author} {\bibfnamefont {D.}~\bibnamefont
  {Chandler}},\ }\href@noop {} {\bibfield  {journal} {\bibinfo  {journal}
  {Proceedings of the National Academy of Sciences}\ }\textbf {\bibinfo
  {volume} {100}},\ \bibinfo {pages} {9710} (\bibinfo {year}
  {2003})}\BibitemShut {NoStop}%
\bibitem [{\citenamefont {Fredrickson}\ and\ \citenamefont
  {Andersen}(1984)}]{fredrickson1984kinetic}%
  \BibitemOpen
  \bibfield  {author} {\bibinfo {author} {\bibfnamefont {G.~H.}\ \bibnamefont
  {Fredrickson}}\ and\ \bibinfo {author} {\bibfnamefont {H.~C.}\ \bibnamefont
  {Andersen}},\ }\href@noop {} {\bibfield  {journal} {\bibinfo  {journal}
  {Physical Review Letters}\ }\textbf {\bibinfo {volume} {53}},\ \bibinfo
  {pages} {1244} (\bibinfo {year} {1984})}\BibitemShut {NoStop}%
  \bibitem [{\citenamefont {Derrida}(1998)}]{derrida1998exactly}%
  \BibitemOpen
  \bibfield  {author} {\bibinfo {author} {\bibfnamefont {B.}~\bibnamefont
  {Derrida}},\ }\href@noop {} {\bibfield  {journal} {\bibinfo  {journal}
  {Physics Reports}\ }\textbf {\bibinfo {volume} {301}},\ \bibinfo {pages} {65}
  (\bibinfo {year} {1998})}\BibitemShut {NoStop}%
  \bibitem [{\citenamefont {Weber}\ \emph {et~al.}(2013)\citenamefont {Weber},
  \citenamefont {Jack},\ and\ \citenamefont {Pande}}]{weber2013emergence}%
  \BibitemOpen
  \bibfield  {author} {\bibinfo {author} {\bibfnamefont {J.~K.}\ \bibnamefont
  {Weber}}, \bibinfo {author} {\bibfnamefont {R.~L.}\ \bibnamefont {Jack}}, \
  and\ \bibinfo {author} {\bibfnamefont {V.~S.}\ \bibnamefont {Pande}},\
  }\href@noop {} {\bibfield  {journal} {\bibinfo  {journal} {Journal of the
  American Chemical Society}\ }\textbf {\bibinfo {volume} {135}},\ \bibinfo
  {pages} {5501} (\bibinfo {year} {2013})}\BibitemShut {NoStop}%
\bibitem [{\citenamefont {Weber}\ \emph {et~al.}(2014)\citenamefont {Weber},
  \citenamefont {Jack}, \citenamefont {Schwantes},\ and\ \citenamefont
  {Pande}}]{weber2014dynamical}%
  \BibitemOpen
  \bibfield  {author} {\bibinfo {author} {\bibfnamefont {J.~K.}\ \bibnamefont
  {Weber}}, \bibinfo {author} {\bibfnamefont {R.~L.}\ \bibnamefont {Jack}},
  \bibinfo {author} {\bibfnamefont {C.~R.}\ \bibnamefont {Schwantes}}, \ and\
  \bibinfo {author} {\bibfnamefont {V.~S.}\ \bibnamefont {Pande}},\ }\href@noop
  {} {\bibfield  {journal} {\bibinfo  {journal} {Biophysical journal}\ }\textbf
  {\bibinfo {volume} {107}},\ \bibinfo {pages} {974} (\bibinfo {year}
  {2014})}\BibitemShut {NoStop}%
  \bibitem [{\citenamefont {Heyl}(2017)}]{heyl2017dynamical}%
  \BibitemOpen
  \bibfield  {author} {\bibinfo {author} {\bibfnamefont {M.}~\bibnamefont
  {Heyl}},\ }\href@noop {} {\bibfield  {journal} {\bibinfo  {journal} {arXiv
  preprint arXiv:1709.07461}\ } (\bibinfo {year} {2017})}\BibitemShut {NoStop}%
\bibitem [{\citenamefont {Cagnetta}\ \emph {et~al.}(2017)\citenamefont
  {Cagnetta}, \citenamefont {Corberi}, \citenamefont {Gonnella},\ and\
  \citenamefont {Suma}}]{cagnetta2017large}%
  \BibitemOpen
  \bibfield  {author} {\bibinfo {author} {\bibfnamefont {F.}~\bibnamefont
  {Cagnetta}}, \bibinfo {author} {\bibfnamefont {F.}~\bibnamefont {Corberi}},
  \bibinfo {author} {\bibfnamefont {G.}~\bibnamefont {Gonnella}}, \ and\
  \bibinfo {author} {\bibfnamefont {A.}~\bibnamefont {Suma}},\ }\href@noop {}
  {\bibfield  {journal} {\bibinfo  {journal} {Physical review letters}\
  }\textbf {\bibinfo {volume} {119}},\ \bibinfo {pages} {158002} (\bibinfo
  {year} {2017})}\BibitemShut {NoStop}%
  \bibitem [{\citenamefont {Gokhale}\ \emph {et~al.}(2014)\citenamefont
  {Gokhale}, \citenamefont {Nagamanasa}, \citenamefont {Ganapathy},\ and\
  \citenamefont {Sood}}]{gokhale2014}%
  \BibitemOpen
  \bibfield  {author} {\bibinfo {author} {\bibfnamefont {S.}~\bibnamefont
  {Gokhale}}, \bibinfo {author} {\bibfnamefont {K.~H.}\ \bibnamefont
  {Nagamanasa}}, \bibinfo {author} {\bibfnamefont {R.}~\bibnamefont
  {Ganapathy}}, \ and\ \bibinfo {author} {\bibfnamefont {A.}~\bibnamefont
  {Sood}},\ }\href@noop {} {\bibfield  {journal} {\bibinfo  {journal} {Nature
  Communications}\ }\textbf {\bibinfo {volume} {5}} (\bibinfo {year}
  {2014})}\BibitemShut {NoStop}%
  \bibitem [{\citenamefont {Keys}\ \emph {et~al.}(2011)\citenamefont {Keys},
  \citenamefont {Hedges}, \citenamefont {Garrahan}, \citenamefont {Glotzer},\
  and\ \citenamefont {Chandler}}]{keys2011}%
   \BibitemOpen
  \bibfield  {author} {\bibinfo {author} {\bibfnamefont {A.~S.}\ \bibnamefont
  {Keys}}, \bibinfo {author} {\bibfnamefont {L.~O.}\ \bibnamefont {Hedges}},
  \bibinfo {author} {\bibfnamefont {J.~P.}\ \bibnamefont {Garrahan}}, \bibinfo
  {author} {\bibfnamefont {S.~C.}\ \bibnamefont {Glotzer}}, \ and\ \bibinfo
  {author} {\bibfnamefont {D.}~\bibnamefont {Chandler}},\ }\href@noop {}
  {\bibfield  {journal} {\bibinfo  {journal} {Physical Review X}\ }\textbf
  {\bibinfo {volume} {1}},\ \bibinfo {pages} {021013} (\bibinfo {year}
  {2011})}\BibitemShut {NoStop}%
\bibitem [{\citenamefont {Chandler}\ and\ \citenamefont
  {Garrahan}(2010)}]{chandler2010dynamics}%
  \BibitemOpen
  \bibfield  {author} {\bibinfo {author} {\bibfnamefont {D.}~\bibnamefont
  {Chandler}}\ and\ \bibinfo {author} {\bibfnamefont {J.~P.}\ \bibnamefont
  {Garrahan}},\ }\href@noop {} {\bibfield  {journal} {\bibinfo  {journal}
  {Annual Review of Physical Chemistry}\ }\textbf {\bibinfo {volume} {61}},\
  \bibinfo {pages} {191} (\bibinfo {year} {2010})}\BibitemShut {NoStop}%
\end{thebibliography}

\end{document}